\documentstyle[aps,prl,epsf,floats]{revtex}

\begin{document}
\twocolumn[{\hsize\textwidth\columnwidth\hsize\csname
@twocolumnfalse\endcsname
\title{
\draft
Effects of Noise in Symmetric Two-Species Competition
}
\author{
Jos\'e M. G. Vilar$^1$ and Ricard V. Sol\'e$^{2}$
}
\address{
$^1$Departament de F\'{\i}sica Fonamental, Facultat de
F\'{\i}sica, Universitat de Barcelona,
Diagonal 647, E-08028 Barcelona, Spain\\
$^2$Department of Physics, FEN, Universitat Polit\'ecnica de Catalunya,
Campus Nord, M\`odul B4,
E-08034 Barcelona, Spain\\
}
\maketitle
\widetext
\begin{abstract}
\leftskip 54.8pt
\rightskip 54.8pt
We have analyzed the interplay between noise and periodic modulations
in a classical Lotka-Volterra model of two-species competition.
We have found that the consideration of noise
changes drastically the behavior of the system and leads to new situations
which have no counterpart in the deterministic case.
Among others, noise is responsible for temporal oscillations, spatial
patterns, and the enhancement of the response of the system
via stochastic resonance.
\end{abstract}
\pacs{PACS numbers: 87.10.+e, 05.40.+j}

}]
\narrowtext

In the last years the study of deterministic mathematical models of
ecosystems has clearly revealed a
large variety of phenomena,
ranging from deterministic chaos to the presence of a spatial
organization \cite{JB1,murray,may3,bale}.
In the absence of spatial dependence these models study the time evolution
of the averaged number of individuals of some interacting species.
When space is considered, the description is usually done in terms of 
population density fields.
These models, however,
do not account for the effects of noise
despite  it is always present in actual population dynamics
and arises from different sources, such as
the intrinsic stochasticity
associated to the dynamics of the individuals and the random variability of
the environment \cite{mont}.
Frequently, its effects have been
assumed to be only a source of disorder\cite{est}.
Conversely, in this Letter we show how the consideration of noise
in a classical Lotka-Volterra model of two species competition
changes drastically and in an unexpected fashion the dynamics of
the deterministic case.
Moreover, consideration of noise and space together leads
to the appearance of spatiotemporal patterns which in the
deterministic model, except for an initial transient and
no matter the value of the parameters, always looks homogeneous.

Let us consider a classical Lotka-Volterra model of symmetric two-species
competition\cite{compe,murray} with the addition of noise terms\cite{mont}
defined by the equations
\begin{eqnarray}
\label{mod11}
{dx \over dt} &=& \mu x ( 1 - x - \beta y) + f_x(x,y)\xi_x(t) \;\; , \\
\label{mod12}
{dy \over dt} &=& \mu y ( 1 - y - \beta x) + f_y(x,y)\xi_y(t) \;\; , 
\end{eqnarray}
where $x$ and $y$ are the population densities, $\mu$ is
proportional to the growth rate, and $\beta$ accounts
for the interactions among the
species. Here the terms $f_i(x,y)\xi_i(t)$ ($i=x,y$)
model the contribution of the random forces.
For the sake of simplicity we assume that $f_x(x,y)=x$,
$f_y(x,y)=y$. Moreover, $\xi_i(t)$ is assumed Gaussian
white noise with zero mean and correlation function
$\left< \xi_i(t) \xi_j(t+\tau) \right> = \sigma\delta(\tau)\delta_{i,j}$
($i,j=x,y$).
This explicit form of the noise term may represent, for instance,
a fluctuating growth rate.
Without the noise terms this Lotka-Volterra model has been widely studied and
it is well know that, for $\beta<1$, both species are present but
for $\beta>1$, exclusion takes place
through  a symmetry-breaking bifurcation and one of the species is 
eliminated.

\begin{figure}[t]
\centerline{
\epsfxsize=5.0cm 
\epsffile{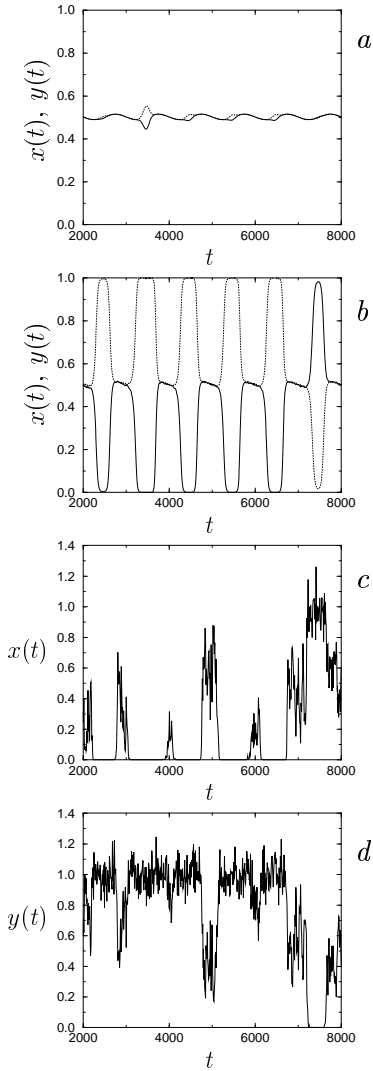}
}
\caption[b]{\label{fig1}
(a) Time evolution of both population densities
[Eqs. (\ref{mod11}) and (\ref{mod12})].
The values of the parameters are $\mu=1$,
$\omega_0/2\pi=10^{-3}$, $\alpha=0.05$,
$\varepsilon=-0.01$, and $\sigma=10^{-12}$.
(b) Same situation as in case (a) but $\sigma=10^{-6}$.
(c) and (d) Same situation as in case (a) but $\sigma=10^{-2}$.
}
\end{figure}

\begin{figure}[t]
\centerline{
\epsfxsize=7cm 
\epsffile{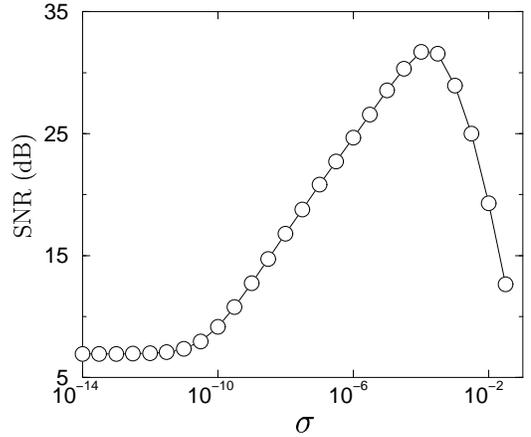}
}
\caption[b]{\label{fig2}
SNR corresponding to Eqs. (\ref{mod11}) and (\ref{mod12}).
The values of the parameters are $\mu=1$,
$\omega_0/2\pi=10^{-3}$, $\alpha=0.05$,
and $\varepsilon=-0.01$
}
\end{figure}

\begin{figure}[t]
\centerline{
\epsfxsize=9cm 
\epsffile{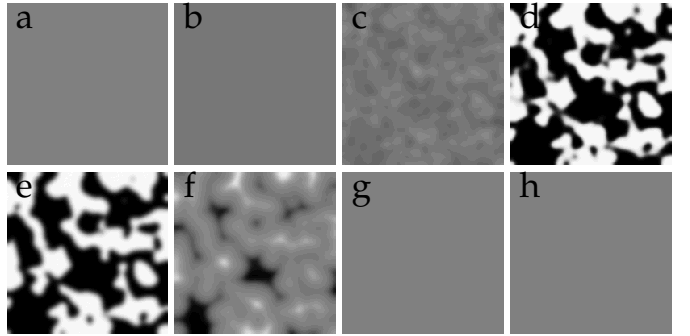}
}
\caption[b]{\label{fig3}
Representation of $x^n_{i,j}$ given
through Eqs. (\ref{mod21}) and (\ref{mod22})
for $n=1750$ (a), $n=1875$ (b), $n=2000$ (c), $n=2125$ (d),
$n=2250$ (e), $n=2375$ (f), $n=2500$ (g), and $n=2625$ (h).
The values of the parameters are $\mu=2$, $D=0.05$,
$\omega_0/2\pi=10^{-3}$, $\alpha=0.1$,
$\varepsilon=-0.01$, and $\sigma=10^{-8}$.
The system size is $200\times200$.
Black and white colors stand for minimum and maximum
values, respectively.
}
\end{figure}

\begin{figure}[t]
\centerline{
\epsfxsize=9cm 
\epsffile{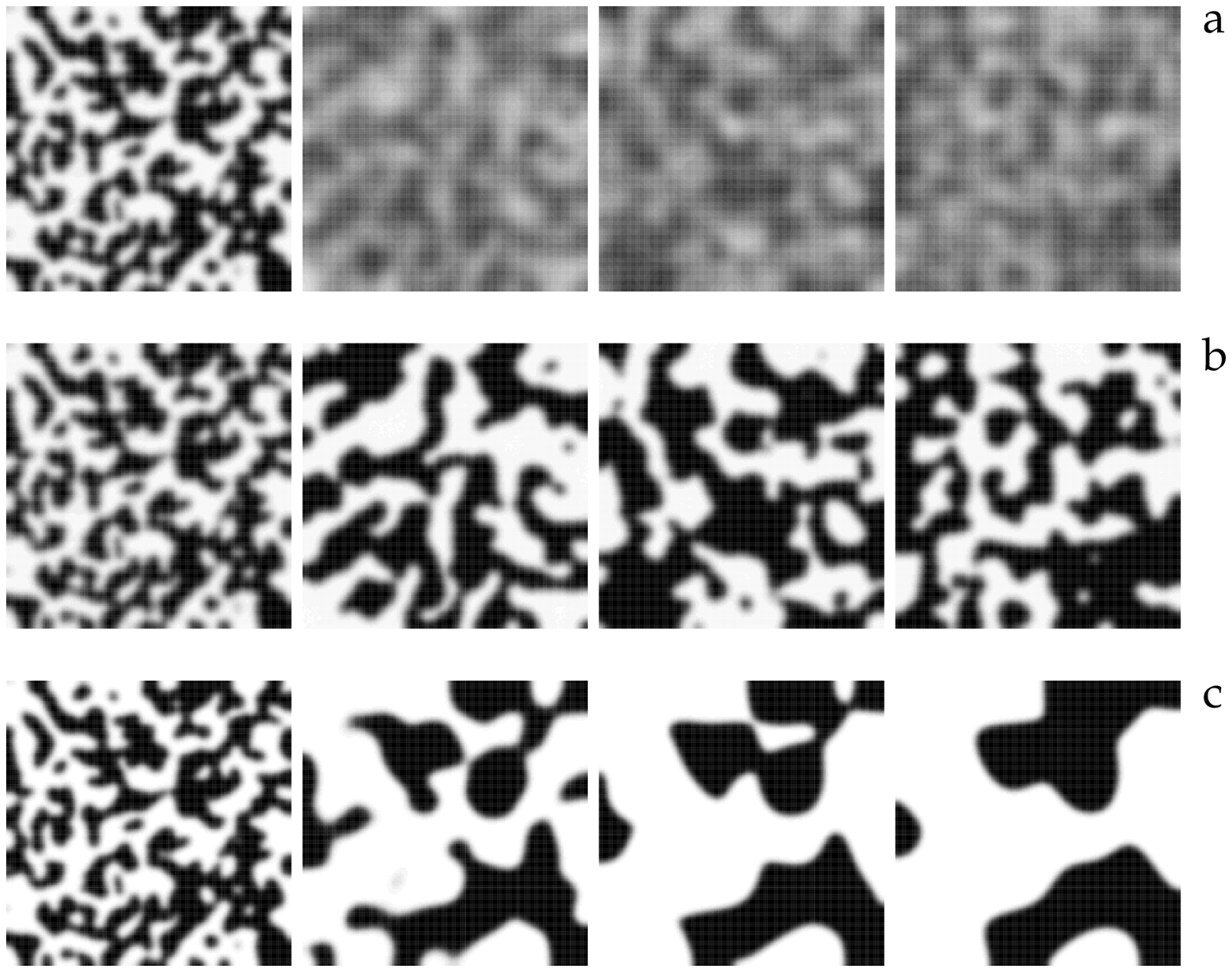}
}
\caption[b]{\label{fig4}
(a) Representation of $x^n_{i,j}$ given through
Eqs. (\ref{mod21}) and (\ref{mod22})
for $n=250, 1250, 2250, 3250$ (from left to right).
The values of the parameters are $\mu=2$, $D=0.05$,
$\omega_0/2\pi=10^{-3}$, $\alpha=0.1$,
$\varepsilon=-0.01$, and $\sigma=10^{-12}$.
The system size is $200\times200$.
Black and white colors stand for minimum and maximum
values, respectively.
(b) Same situation as in case (a) but $\sigma=10^{-8}$.
(c) Same situation as in case (a) but $\sigma=0$ and $\varepsilon=0.01$.
}
\end{figure}

As an additional feature, we account for a non-stationary environment.
The simplest way to do it is to assume that our dynamics is modulated
through a periodic variation in one of the relevant parameters.
Seasonality and daily variations are the most important sources
of regular forcing. 
The explicit case we consider here is described by
\begin{equation}
\beta\equiv\beta(t)=1+\varepsilon+\alpha \sin(\omega_0 t) \;\; ,
\end{equation}
where $\varepsilon$, $\alpha$ and $\omega_0$ are constants.
This form of $\beta(t)$ corresponds to a situation in which
the competition between
species is altered in such a way that we move from
coexistence ($\beta < 1$) to exclusion ($\beta >1$) in a regular fashion.
The change in the competition rate may
occur when a limiting resource which is used by both populations
goes from large to low values.
Thus, at high levels, competition is weak, but can become
strong at low levels.

Although some Lotka-Volterra models with sinusoidal perturbations of
parameters show very complex dynamical patterns\cite{kut},
in the model we consider the situation
is rather simple. In the absence of noise,
except for a possible initial transient,
when $\varepsilon>0$
exclusion takes place.
On the other hand, when $\varepsilon<0$ 
the species always coexist, in spite of the fact that $\beta$
becomes larger than $1$ periodically in time.
The introduction of noise drastically changes the previous scenario since
even small amounts of stochasticity are able 
to destroy the state of 
coexistence periodically\cite{caca}.  In order to analyze our system
we have numerically integrated the previous equations by
using standard methods for stochastic differential equations \cite{kloeden}.
In Fig. \ref{fig1}a
we have plotted the temporal evolution of both species
when $\varepsilon<0$ and the noise level $\sigma$ is very small,
but its effects
are slightly appreciable. Thus, the species  practically
coexist all the time.
When noise increases sufficiently a species is able to nearly exclude
the other one periodically (see Fig. \ref{fig1}b), i.e., one species dominates
the other. 
It is interesting to realize that by increasing even more the noise level
this coherent response to
environmental variations is lost (See Figs. 1c and 1d).
A nontrivial consequence of these results is that time variations
in competing populations
could be a noise-induced phenomenon. In this context,
periodic population oscillations, not present
in the original deterministic model, are now allowed to appear.

The previous figure exhibits an aspect which has 
received a considerable attention in the recent years; namely, the response
of the system to a periodic force may be
enhanced by the presence of noise. This 
is the main characteristic of stochastic resonance (SR)\cite{SR1a,SR1b,SR1c}.
In essence,
SR is a nonlinear cooperative effect in which a weak periodic
stimulus entrains large-scale environmental
changes. These changes are done in a coherent fashion and, as a result,
the periodic component is greatly enhanced.
The usual form to make the presence of SR manifest is through the
signal-to-noise ratio (SNR) \cite{SR1a,nueva} of a given quantity describing
the state of the system. Thus, SR arises when the SNR has a maximum as
a function of the noise level. 

In the system we consider
the most straightforward form of measuring qualitative changes
is by using a quantity that accounts for the degree
of coexistence, such as the squared difference of population
densities $(x-y)^2$.
In Fig. \ref{fig2} we have shown the SNR for this quantity which clearly
exhibits a maximum at non-zero noise level,
thus, indicating the presence of SR\cite{mult}.
In fact, this is the first example of SR in a population dynamics model.
In biology these situations have been restricted primarily to
physiological systems\cite{SR2}, but other areas, as population dynamics, have
not been explored up to now.
We should note that previously studied systems displaying SR
always exhibit oscillations when noise is absent by
only changing the values of the parameters, usually
by increasing the amplitude of the input signal.
Conversely, in the case we are concerned oscillations
only appears when noise is present.

A further step in our study will be the analysis of
the effects of space.
The usual way to do it is by adding diffusive terms to
Eqs. (\ref{mod11}) and (\ref{mod12}).
This class of spatiotemporal model corresponds to the situation in which
the dynamics of the species is continuous in time.
Other situations of interest concerns with the description of
populations  whose generations do not overlap in time\cite{Mayx}.
In this case, the continuous time-space description is no
longer valid and discrete time evolution models must be considered.
The usual way to model spatially distributed systems whose
time evolution is discrete is by using a coupled map lattice (CML)\cite{ka}.
Here we
follow this approach.
In such a situation the dynamics of our discrete model is
\begin{eqnarray}
\label{mod21} x^{n+1}_{i,j}&=&\mu x^{n}_{i,j} ( 1 - x^{n}_{i,j}
 - \beta^{n}y^{n}_{i,j})
\nonumber \\ & & + \sqrt{\sigma}x^{n}_{i,j}{X}^{n}_{i,j}
+D\sum_\gamma(x^{n}_\gamma-x^{n}_{i,j}) \;\; ,\\
\label{mod22} y^{n+1}_{i,j}&=&\mu y^{n}_{i,j} ( 1 - y^{n}_{i,j}
 - \beta^{n}x^{n}_{i,j})
+ \sqrt{\sigma}y^{n}_{i,j}{Y}^{n}_{i,j}
\nonumber \\ & & +D\sum_\gamma(y^{n}_\gamma-y^{n}_{i,j}) \;\; ,
\end{eqnarray}
where $\beta^{n}=1+\varepsilon +\alpha\cos(\omega_0n)$,
$D$ is a constant accounting for the diffusion, and
$\sum_\gamma$ indicates sum over the four nearest neighbors.
Here the random terms are modeled by independent Gaussian
random variables, denoted by ${X}^{n}_{i,j}$ and ${Y}^{n}_{i,j}$,
with zero mean and variance unit.
The remaining parameters have the same meaning as in the
zero dimensional time-continuous model [Eqs. (\ref{mod11}) and (\ref{mod12})].
When $\beta^{n}\equiv 0$ this CML model reduces to a logistic lattice, which
has been widely explored\cite{logCML}.

As in the previous case,  when $\beta^{n} < 1$
this model exhibits a state in which both species
coexist. The coexistence state is responsible for the
appearance of an homogeneous spatial
distribution. However, when a sufficiently large amount of noise
is present and $\beta^{n} > 1$ spatial patterns arise.
In order to elucidate how this spatial structure emerges, we have depicted
in Fig. \ref{fig3} the spatiotemporal evolution corresponding to the
CML model for one period of $\beta^{n}$.
The other periods also exhibit the same characteristics.
Noise is responsible for the periodically appearance of the spatial
structure since 
if low enough, except for an initial transient, 
the system always looks homogeneous.
An example of how patterns are influenced by the noise intensity
is shown in Figs. \ref{fig4}a and \ref{fig4}b. 
We have plotted four spatial patterns corresponding to the 
first four periods of $\beta^{n}$ for
two representative values of the noise level.
The first pattern is strongly influenced by the initial conditions,
which are random, and looks very similar for each noise intensity.
However, once the
initial transient is lost, the patterns corresponding to the 
higher noise level are more pronounced than the ones 
with lower noise level. If noise is sufficiently decreased 
patterns do not appear.
A similar behavior is also present in the continuous
time-space model.

It is worth emphasizing that
the deterministic counterpart of most models
exhibiting noise induced structures is able
to display  similar patterns to those induced by noise for a certain
range of the values of the parameters, as for instance
the Swift-Hohenberg equation \cite{prl3}.
In our model, however,
these patterns only arise when noise is present.
For instance if $\varepsilon > 0$ a spatial structure
emerges periodically over a long transient, but eventually it
disappears. In Fig. \ref{fig4}c we have depicted four patterns
corresponding to this situation. This figure clearly shows
how the domains grow in each period. At suffiently large
time there is one domain, thereby one species excludes the other.

In summary, we have shown that noise cannot systematically be
neglected in models of population dynamics.
Its presence is responsible for the generation
of temporal oscillations and for the appearance
of spatial patterns. In contrast, these features
do not arise when noise is absent.
Under some circumstances noise has a 
constructive role, since it is responsible
for the enhancement of the response of the system
via stochastic resonance.
The similarity of the models we have considered
with phase separation, has been already
pointed out in the literature \cite{kap,sm,onu}.
In this regard, it is worth emphasizing that
there exist numerous systems which can be
described through competitive or cooperative interactions.
To mention just a few: biological assemblies of individuals,
coupled chemical reactions, political parties, business,
and countries \cite{mont}.
Thus, our results are not restricted
only to population dynamics, but the main ideas can be applied to a wide
variety of situations
embracing different scientific areas.

The authors would like to thank J. M. Rub\'{\i} for
fruitful discussions.
This work was
supported by DGICYT of the Spanish Government under Grants Nos.
PB95-0881 and PB94-1195.
J. M. G. V. wishes to thank Generalitat
de Catalunya for financial support.


\begin{references}
\bibitem{compe} This is one of the most celebrated models in theoretical
ecology and describes the interaction of two given species which share
a common resource.
\bibitem{caca} When $\varepsilon>0$ noise does not change in a qualitative
fashion the behavior of the system because one
species is also excluded.
\bibitem{nueva} The effect of a periodic force may
be analyzed by means of the power spectrum
$P(\omega)=\int_0^{2\pi/\omega_0}dt \int_{-\infty}^\infty 
\left<v(t)v(t+\tau)\right>e^{-i\omega \tau}d\tau $,
where $v(t)$ is a quantity describing the state of the system.
It consists of a delta function
centered at the frequency $\omega_0$ plus a function $Q(\omega)$
which is smooth in the neighborhood of $\omega_0$ and is given by
$P(\omega) %\equiv \vert \hat v(\omega) \vert^2
=Q(\omega)+S(\omega_0)\delta(\omega-\omega_0)$.  Hence, the SNR of
$v(t)$ is defined by ${\rm SNR} \equiv {S(\omega_0)/Q(\omega_0)}$.
\bibitem{mult}
We would like emphasize, that at low noise levels the form of the 
noise term is unimportant for the dynamics of the system.
For higher noise level,
due to the multiplicative character of the noise, the dynamics of the
system depend of the explicit form of the noise term. However, the
presence of SR is also manifested.
\end{references}
\end{document}